\documentclass[titlepage,12pt]{article}
\usepackage{textcomp}
\usepackage{chetnew}
\usepackage{amssymb,amsmath,color,graphics,amscd,epsf,indentfirst,amsfonts}
\usepackage{mathtools}
\usepackage{enumerate}
\usepackage{epsfig}
\usepackage{slashed}

\usepackage{tikz}

\def\blfootnote{\xdef\@thefnmark{}\@footnotetext}

\long\def\symbolfootnote[#1]#2{\begingroup%
\def\thefootnote{\fnsymbol{footnote}}\footnote[#1]{#2}\endgroup}

\makeatletter
\renewcommand{\@dotsep}{4.5}
\makeatother

\def\be{\begin{equation}}
\def\ee{\end{equation}}

\makeatletter
\def\@seccntformat#1{\csname the#1\endcsname.\quad}
\makeatother

\def\clock{{\count0=\time
           \divide\count0 60
           \ifnum\count0<10 0\fi\the\count0
           \multiply\count0 -60 \advance\count0 \time
           :\ifnum\count0<10 0\fi \the\count0
         }}
\newcommand{\timestamp}{{\small\vbox{\hbox{\tt\jobname.tex}
\hbox{\the\day/\the\month/\the\year, \clock}}}}

\def\ZZ{{\cal Z}}

\def\Zk{{\mathbb{Z}_k}}

\def\time{{\tau}}

\def\beq{\begin{equation}}
\def\eeq{\end{equation}}
\newcommand{\bea}{\begin{eqnarray}}
\newcommand{\eea}{\end{eqnarray}}
\def\bal{\begin{align}}
\def\eal{\end{align}}

\def\drawbox#1#2{\hrule height#2pt
         \hbox{\vrule width#2pt height#1pt \kern#1pt
               \vrule width#2pt}
               \hrule height#2pt}

\def\Asym#1#2{\vcenter{\vbox{\drawbox{#1}{#2}
               \kern-#2pt       
               \drawbox{#1}{#2}}}}

\numberwithin{equation}{section}

\begin{document}
\begin{titlepage} 
\vskip 4cm
\begin{center}
\font\titlerm=cmr10 scaled\magstep4
    \font\titlei=cmmi10 scaled\magstep4
    \font\titleis=cmmi7 scaled\magstep4
    \centerline{\LARGE \titlerm 
      S-Dual of Maxwell Chern-Simons Theory}
          \vskip 0.3cm
\vskip 1cm
       {Adi Armoni}\\
\vskip 0.5cm
       {\it Department of Physics, Faculty of Science and Engineering}\\
       {\it Swansea University, SA2 8PP, UK}\\       

\vskip 0.5cm
{a.armoni@swansea.ac.uk}\\

\end{center}
\vskip .5cm
\centerline{\bf Abstract}
\baselineskip 20pt
%

\vskip .5cm 
\noindent
We discuss the dynamics of three dimensional Maxwell theory coupled to a level $k$ Chern-Simons term. Motivated by S-duality in string theory we argue that the theory admits an S-dual description. The S-dual theory contains a non-gauge 1-form field, previously proposed by Deser and Jackiw \cite{Deser:1984kw} and a level $k$ $U(1)$ Chern-Simons term, $\ZZ_{\rm MCS}=\ZZ_{\rm DJ}\ZZ_{\rm CS}$. The couplings to external electric and magnetic currents and their string theory realisations are also discussed.

\vfill
\noindent
\end{titlepage}\vfill\eject

\hypersetup{pageanchor=true}

\setcounter{equation}{0}

\pagestyle{empty}
\small
\normalsize
\pagestyle{plain}
\setcounter{page}{1}
 
\section{Introduction} 
\label{intro}
Chern-Simons theory is vastly used in mathematical physics, in condensed matter physics and in string theory \cite{Dunne:1998qy}. It was studied intensively in the past three decades, yet the dynamics of Yang-Mills Chern-Simons theory is not fully understood at the strong coupling regime.

Four dimensional S-duality is an exact duality between two ${\cal N}=4$ super Yang-Mills theories, enabling us to calculate quantities in the strong coupling regime using a dual weakly coupled theory. In the Abelian case it reduces to the old electric-magnetic duality which swaps electric and magnetic fields
\beq
F \longleftrightarrow *F \, .
\eeq

In 3d Abelian S-duality relates the electric field to a dual scalar
\beq
f \longleftrightarrow * d \phi \, .
\eeq

The purpose of this note is to extend S-duality to 3d Maxwell Chern-Simons (MCS) theory, with either a compact or non-compact $U(1)$ gauge group. It should hold on any spin manifold. The Lagrangian of the theory is given by
\beq
L = -\frac{1}{2g^2} da_e \wedge * da_e + \frac{k}{4\pi} a_e \wedge da_e \, . \label{mcs}
\eeq
MCS theory contains a vector boson of mass $M=\frac{g^2 k}{2\pi}$. At low energies the kinetic term is irrelevant and the theory flows to a pure level $k$ Chern-Simons theory. As explained in section \eqref{section-zk} the theory admits a global $\mathbb{Z}_k$ 1-form symmetry generated by
\beq
 G \equiv \exp \left ( i  \oint  ( a_e - \frac{1}{M} *da_e) \right ) \, .
 \eeq
When the theory is compactified on a torus, the global $\Zk$ 1-form symmetry is  spontaneously broken, resulting in $k$ degenerate vacua.
 
Several attempts were made to find the S-dual of \eqref{mcs}. In \cite{Deser:1984kw} Deser and Jackiw proposed a 'self-dual model' (SDM) which describes a massive vector. While SDM describes a massive vector it does not admit a $\Zk$ 1-form symmetry neither does it flow to a pure Chern-Simons theory at low energies, hence it cannot be an exact dual of MCS theory.

\begin{figure}[!th]
\centerline{\includegraphics[width=9cm]{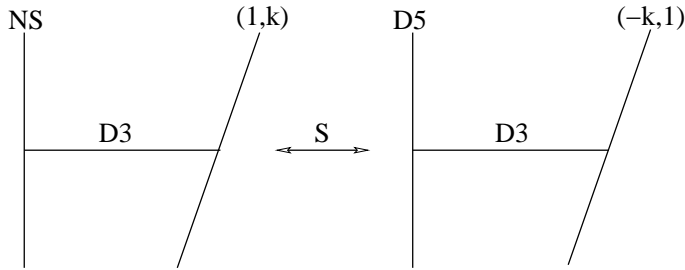}}
\caption{\footnotesize The electric theory on the left brane configuration is Maxwell Chern-Simons. The magnetic theory, obtained by type IIB S-duality, lives on the right brane configuration.}
\label{branes}
\end{figure}

A closely related problem concerns the open string dynamics on a certain Hanany-Witten brane configuration. It is well known \cite{Kitao:1999uj,Bergman:1999na} that MCS theory lives on the left brane configuration of fig.\eqref{branes}. Type IIB S-duality maps the left configuration into the right configuration. Thus, knowing the field theory that lives on the right configuration will solve the problem of finding the S-dual. In early attempts \cite{Kitao:1999uj,Bergman:1999na,Kapustin:1999ha} the authors found gauge theories with a fractional level Chern-Simons term. While the theories they found are classically equivalent to MCS, it cannot be the full answer, as it does not admit the symmetries nor the same dynamics as the electric theory.

\section{Derivation of the duality}

We may use 4d S-duality between Maxwell theories to derive the 3d duality.
In 4d a pure Maxwell theory with a coupling $g$ is dual to a pure Maxwell theory with a coupling $1/g$.

Consider the following partition function
\beq
\ZZ=\int DF_m DA_e \exp i\int (-\frac{g^2}{2} F_m \wedge *F_m +  F_m \wedge dA_e) \label{master}
\eeq
$A_e$ is the 'electric' gauge field, $F_m$ is a 'magnetic' gauge invariant 2-form. $g$ is the 'electric' gauge coupling.

Upon integrating over $F_m$ we obtain the electric theory
\beq
\ZZ=\int DA_e \exp i\int \left ( -\frac{1}{2g^2} dA_e \wedge *dA_e \right ) \, .
\eeq

If instead we integrate over $A_e$ we obtain
\beq
\ZZ=\int DF_m \delta (dF_m) \exp i\int \left ( -\frac{g^2}{2} F_m \wedge *F_m \right ) \, ,
\eeq
hence it can be written in terms of $A_m$ such that $F_m=dA_m$.
\beq
\ZZ=\int DA_m \exp i \left ( -\frac{g^2}{2} dA_m \wedge *dA_m \right ) \, .
\eeq
 This is the magnetic theory dual to the electric theory.

 Let us use dimensional reduction of \eqref{master} in order to derive the 3d duality. Upon reducing to 3d the 4d 2-form $F_m$ becomes a 3d 2-form $f_m$ and 1-form $a_m$. The 4d gauge field $A_e$ becomes a 3d gauge field $a_e$ and a scalar $\phi_e$. The the 2-form $f_m$ and the scalar $\phi_e$ decouple from the rest of the action and admit
 \beq
 \ZZ = \int Df_m D\phi_e \exp i\int  (-\frac{g^2}{2} f_m \wedge *f_m + f_m \wedge d\phi _e )  \, ,
 \eeq
 which leads to the well known S-duality
 \beq
 d\hat a _m\longleftrightarrow * d \phi _e \, ,
\eeq
where $f_m=d\hat a _m$.

 Let us focus on the duality between $a_e$ and $a_m$, which is the prime purpose of this note. We add to the action a Chern-Simons term\footnote{The Chern-Simons term can be obtained by a dimensional reduction of a space dependent theta term $\int d^4 x \, \theta(x) F_e \wedge F_e$, such that $\theta (x) =\frac{k}{4\pi} H(x^3)$, where $H(x^3)$ is the Heaviside step function.}. Our proposal is the following 'master' partition function
\beq
\ZZ=\int Da_m Da_e \exp i\int (-\frac{g^2}{2} a_m \wedge *a_m + a_m \wedge da_e + \frac{k}{4\pi} a_e \wedge da_e ) \label{master-3d}
\eeq
Note that $a_m$ is a {\it gauge invariant} 1-form. Upon integration over $a_m$ we obtain the electric theory
\beq
\ZZ=\int Da_e \exp i\int (-\frac{1}{2g^2} da_e \wedge *da_e + \frac{k}{4\pi} a_e \wedge da_e )  \, , \label{electric}
\eeq
namely Maxwell Chern-Simons theory.

In order to derive the magnetic theory we should use \eqref{master-3d} and integrate over $a_e$. This is a subtle point. Instead, let us use a change of variables $a_e = b - \frac{2\pi}{k}a_m$, to obtain the following partition function
\beq
\ZZ=\int Da_m Db \exp i\int (-\frac{g^2}{2} a_m \wedge * a_m - \frac{\pi}{k} a_m \wedge da_m + \frac{k}{4\pi} b \wedge db ) . \label{magnetic}
\eeq

Equation \eqref{magnetic} is our proposal for the S-dual of Maxwell Chern-Simons theory. The partition function of the magnetic theory is a product of the Deser-Jackiw theory and a level $k$ Chern-Simons term
\beq
\ZZ_{\rm MCS} = \ZZ_{\rm DJ}\ZZ_{\rm CS} \, .
\eeq
Note that $a_m$ is not a gauge field and therefore the term $\frac{\pi}{k} a_m \wedge da_m$ is not ill-defined.

Both the electric and the magnetic theories describe a massive vector of mass $M= \frac{g^2 k}{2\pi}$ and a decoupled level $k$ Chern-Simons theory. Both theories exhibit a 1-form $\Zk$ symmetry.

Let us now provide another argument in favour of our proposal \eqref{magnetic}. We begin with the magnetic brane configuration of fig.\eqref{branes}. It was argued by Gaiotto and Witten \cite{Gaiotto:2008ak} that the theory which lives on the intersection of the 3-brane and the tilted 5-brane (without a D5 brane) is
\beq
\ZZ= \int Da Dc \exp i \int (\frac{1}{2\pi} a \wedge dc + \frac{k}{4\pi} c \wedge dc) \, . \label{gw}
\eeq
In order to understand what happens when we add a D5 brane, let us assume that the terms that we need to add to the action are $k$ independent. Indeed, the information about $k$ is encoded in the tilted fivebrane, not in the threebrane. Let us use $k=0$, since in this case the duality is well understood: the electric theory is pure Maxwell and the magnetic (mirror) theory is a massless scalar. The brane realisation of the duality was provided in the seminal work of Hanany and Witten \cite{Hanany:1996ie}.

We may write the theory of a free massless scalar as follows 
\beq
\ZZ= \int Da Dc \exp i \int (a \wedge *a + \frac{1}{2\pi} a \wedge dc) \, ,
\eeq
with $a$ being a gauge invariant 1-form.  
The equation of motion for $c$ is $*da=0$, namely that $a=d \chi$. Thus, for $k=0$ we obtain a theory of a free scalar $(d\chi)^2$, as expected. 

We found that adding a term $a \wedge *a$ to the action yields a theory that describes the correct dual of Maxwell theory. We propose that $a \wedge *a$ is the missing term in \eqref{gw}, namely that by adding it to \eqref{gw} we obtain the dual of MCS for any $k$. Note that
\beq
\ZZ= \int Da Dc \exp i \int (a \wedge *a + \frac{1}{2\pi} a \wedge dc + \frac{k}{4\pi} c \wedge dc) \, \label{master2} 
\eeq
is almost identical to \eqref{master-3d}. An important difference is that Gaitto and Witten introduced a {\it gauge field} $a$, whereas in \eqref{master2} we added a term that breaks gauge invariance. We may re-introduce gauge invariance in \eqref{master2} by transforming the fixed gauge vector $a$ into a gauge invariant term by adding a scalar $\eta$ as follows
\beq
\ZZ= \int Da Dc D\eta \exp i \int ((a -d\eta) \wedge *(a -d\eta) + \frac{1}{2\pi} a \wedge dc + \frac{k}{4\pi} c \wedge dc) \, \label{master3} \, ,
\eeq
such that under a gauge transformation $a \rightarrow a +d\lambda, \eta \rightarrow \eta + \lambda$, with $a$ a $U(1)$ gauge field. Eq.\eqref{master2} may be viewed as the fixed gauge version of eq.\eqref{master3} with $d\eta =0$. 

Our proposal \eqref{magnetic} passes all the requirements from a dual theory: it admits a $\Zk$ global symmetry, it flows to pure Chern-Simons theory in the IR, it contains a massive vector of mass $M$ and, finally, when $k=0$ it agrees with the results of Hanany and Witten \cite{Hanany:1996ie}. As we shall see the brane realisations of both electric and magnetic theories predict the existence of $k$ degenerate vacua.

We summarize this section by writing the precise map between the electric and magnetic variables using \eqref{master-3d}
\bea
-g^2 a_m = *da_e  \\
 b = a_e -\frac{1}{M} *da_e 
\eea
or
\beq
a_e = b- \frac{2\pi}{k} a_m
\eeq
 
 \section{Comments on $\Zk$}
\label{section-zk}

 Let us introduce a Wilson loop in MCS theory. We wish to measure the $\Zk$ charge of the loop, namely the number of fundamental strings, $n$, that pass through a a certain contour $C$. We will define an operator $G$ such that
 \beq
 G W_n = \exp (i\frac{2\pi n}{k}) W_n \, ,
 \eeq
 with $W_n$ a Wilson loop of charge $n$, $W_n = \exp (i n \oint a_e)$.
 
\begin{figure}[!th]
\centerline{\includegraphics[width=4cm]{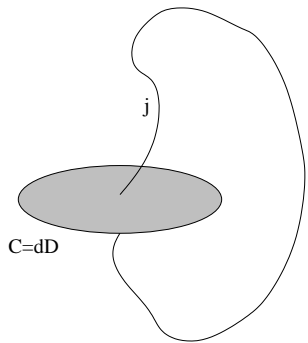}}
\caption{\footnotesize A Wilson loop passing through a domain $D$ (shaded region) whose boundary is $C=\partial D$.}
\label{zk}
\end{figure}

In order to the define $G$, let us consider the equation of motion in MCS
 \beq
 d(\frac{1}{g^2} *da_e - \frac{k}{2\pi} a_e) = j_e \equiv dJ_e \, , \label{eom}
 \eeq
 where $J_e$ is the integral of the electric current $j_e$ over a disc $D$ such that $C=\partial D$. The setup is depicted in fig.\eqref{zk}. By integrating \eqref{eom} we learn that
 \beq
  \frac{1}{g^2} *da_e - \frac{k}{2\pi} a_e = J_e \, .
  \eeq

 We can therefore define a generator of a $\Zk$ symmetry as follows 
\beq
 G= \exp \left ( i \frac{2\pi }{k} \oint _C (\frac{k}{2\pi} a_e - \frac{1}{g^2} *da_e) \right ) = \exp \left ( i  \oint _C  b \right )  .
 \eeq

 Note that the implication of $\Zk$ symmetry is a symmetry $n \rightarrow n +k$, namely that a collection of $k$ strings is topologically isomorphic to a singlet, namely to no strings at all. This is supported by string theory: suppose that we attempt to place the endpoints of $k$ coincident strings on the worldvolume of the D3 brane. The collection of $k$ fundamental strings can transform itself into an anti D-string and a $(k,1)$ string. Instead of ending on the worldvolume of the D3 brane, the D-string can end on an NS5 brane and a $(k,1)$ string can end on a $(1,k)$ fivebrane. Thus string theory predicts that a collection of $k$ strings can be removed from the worldvolume of the 3d gauge theory. A similar phenomenon happens in the magnetic dual if we attempt to introduce $k$ coincident D-strings in the worldvolume of the magnetic theory.

 When the theory is defined on the torus the $\Zk$ symmetry is broken, resulting in $k$ vacua \cite{Ho:1992dy}\cite{Tong}.  An intuitive explanation is as follows: The level $k$ $U(1)$ Chern-Simons theory is equivalent (using level-rank duality) to a level $1$ $SU(k)$ theory, that admits a $\Zk$ centre symmetry. When it is defined on the torus the $SU(k)$ theory deconfines, resulting in $k$ degenerate vacua, parameterized by the eigenvalues of the 't Hooft loop.

 The $k$ vacua manifests themselves in both the electric and magnetic brane configurations as follows: the D3 brane may end on any of the $k$ constituents of the fivebranes. Each one of the $k$ choices corresponds to a vacuum.

 \section{Coupling to external sources, Wilson and magnetic loops} \label{section-coupling}

Consider the coupling of the electric gauge field to a source $j_e$, namely $a_e j_e$. It translates to the coupling $(b- \frac{\pi}{k} a_m)j_e$ in the magnetic side. We therefore suggest that the Wilson loop
 \beq
 W_e=\exp i \oint a_e
 \eeq
in the electric side is mapped into a magnetic loop of the form
  \beq
M_m= \exp i \oint (b- \frac{2\pi}{k} a_m)
 \eeq
 in the magnetic side.

 We may use the above map between the Wilson loop in the electric side and its magnetic counterpart to study the dynamics of 3d QED-CS. Using the worldline formalism \cite{Strassler:1992zr} we can write the partition function of MCS theory coupled to $N_f$ massless fields as follows
 \beq
 \ZZ _{\rm QED-CS} = \int Da_e \exp (iS_{\rm MCS}) \sum_n \frac{(N_f \Gamma_e)^n}{n!}
 \eeq
 with
 \beq
 \Gamma _e = \int \frac{dt}{t^{\frac{5}{2}}}\int Dx \exp (-\int _0 ^t d\tau (\dot x)^2) \exp i \oint a_e 
 \eeq
 The duality transformation yields the following partition function
  \beq
 \ZZ _{\rm magnetic} = \int Da_m Db \exp (iS_{\rm DJ-CS}) \sum_n \frac{(N_f \Gamma_m)^n}{n!}
 \eeq
 with
 \beq
 \Gamma _m = \int \frac{dt}{t^{\frac{5}{2}}}\int Dx \exp (-\int d\tau (\dot x)^2)
  \exp i \oint (b- \frac{2\pi}{k} a_m) \, . \label{gamma_m}
 \eeq
 It suggests that the dynamics of QED with $N_f$ massless flavours is captured by a dual DJ-CS theory coupled to $N_f$ massless 'monopoles'. The precise coupling of $a_m$ and $b$ to the monopoles is given by \eqref{gamma_m}. We may write the dual magnetic theory in a more 'standard' form
  \beq
 \ZZ =\int Da_m Db D\bar \psi _m D\psi _m \exp i S_{\rm magnetic} \, ,
 \eeq
 with $S_{magnetic}$ given by
 \beq
S_{magnetic}= \int \left (-\frac{g^2}{2} a_m \wedge * a_m - \frac{\pi}{k} a_m \wedge da_m + \frac{k}{4\pi} b \wedge db + \bar \psi _m \gamma  \wedge \star (i\partial + b- \frac{2\pi}{k} a_m ) \psi _m \right ) \, .
\eeq
It is interesting to note that the QED-CS theory is mapped to a theory of interacting massless magnetic 'monopoles', with a coupling $1/gk$. Thus, when the electrons couple strongly to $a_e$, the 'monopoles' couple weakly to $a_m$ and we may use perturbation in theory in the magnetic side to study the strongly coupled electric theory.

Following Itzhaki\cite{Itzhaki:2002rc} let us define a magnetic (``disorder'') loop in the electric theory
\beq
 M_e= \exp \left ( i \oint _C ( k a_e - \frac{2\pi }{g^2} *da_e) \right )\, , \label{Me}
 \eeq
 which is mapped into the electric loop in the magnetic side 
\beq
 W_m= \exp \left ( i k \oint _C  b \right ) . \label{Em} 
\eeq 
 The magnetic loop in the electric side and the electric (Wilson) loop in the magnetic side are trivial \cite{Itzhaki:2002rc}. 

 \begin{figure}[!th]
\centerline{\includegraphics[width=9cm]{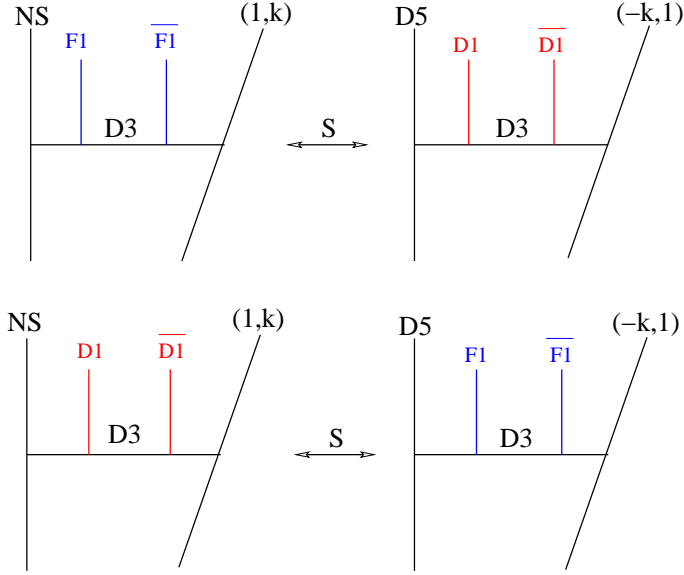}}
\caption{\footnotesize Rectangular Wilson loops can be realised in string theory by ending a pair of F-string and anti F-string on the the threebrane. Similarly Rectangular 't Hooft loops can be realised by ending a pair of D-string and anti D-string on the three brane. The end of the F-string represents a heavy quark, whereas the end of the D-string represents a heavy monopole.}

\label{loops}
 \end{figure}

 We suggest that a rectangular Wilson loop (or magnetic loop) should be identified with the end points of an F-string and anti F-string (or a D-string and anti D-string) that end on the threebranes of fig.\eqref{loops}.

 A D-string can end on an NS5 brane instead of a threebrane, hence the magnetic loop in the electric theory should be trivial. Similarly, a F-string can end on a D5 brane instead of a threebrane, hence a Wilson loop in the magnetic theory should be trivial. This is consistent with our definitions of the magnetic loop \eqref{Me} and the Wilson loop \eqref{Em}. 

\section{Summary}

 The purpose of this note is to find the S-dual of MCS theory. We found that the dual theory \eqref{magnetic} contains a non-gauge vector of mass $M$ and a decoupled pure TQFT. The magnetic theory nicely captures the dynamics of the electric theory: a theory with a mass gap that flows in the IR to a TQFT. The duality we uncovered in this note is a precise manifestation of the duality between a topological insulator and a topological superconductor outlined in ref.\cite{Murugan:2016zal}.

 It will be interesting to find the S-dual of the non-Abelian $U(N)$ theory that lives on a collection of $N$ coincident $D3$ branes, suspended between tilted fivebranes. The master field of that theory may be obtained by replacing the Abelian 1-forms of \eqref{master-3d} by non-Abelian 1-forms as follows\footnote{I thank Shigeki Sugimoto for suggesting that.}
\beq
\ZZ=\int Da_m Da_e \exp i\, {\rm tr} \int (-\frac{g^2}{2} a_m \wedge *a_m + a_m \wedge (da_e + a_e \wedge a_e) + \frac{k}{4\pi} (a_e \wedge da_e+ \frac{2}{3} a_e \wedge a_e \wedge a_e) ) 
\eeq
together with  $a_e = b - \frac{2\pi}{k}a_m$. Other dualities that involve $SO/Sp$ (and an orientifold in string theory) could also be derived. The generalisation to supersymmetric QED/QCD theories with a CS term \cite{Kapustin:1999ha} is also interesting and can be written down using the worldline formalism, as in section \eqref{section-coupling}. The duality found in this letter is useful to study the strong coupling regime of those theories.

 Finally, it is well known that MCS theory admits Seiberg duality. The manifestation of the duality using an exchange of fivebranes in the magnetic theory, might teach us about fivebranes dynamics.
  
\subsection*{Acknowledgements}
A.A. would like to thank the Yukawa institute for theoretical physics, where part of this work was done, for warm hospitality. I would also like to thank Mohammad Akhond and Shigeki Sugimoto for numerous discussions and collaboration. I am grateful to Anton Kapustin and Zohar Komargodski for reading and commenting on a draft version of this paper.


\newpage
\providecommand{\href}[2]{#2}

\end{document}